\documentclass[epj]{svjour}
\usepackage{graphics}

\begin{document}
\titlerunning{
Phase diagram of the Mott Transition in a two-band  Hubbard model
}
\title{
Phase diagram of the Mott Transition in a two-band  Hubbard model in infinite dimensions
}
\author{Y. \=Ono\inst{1,3}, R. Bulla\inst{2} \and A. C. Hewson\inst{3}
}                     
%
%
\institute{Department of Physics, Nagoya University, Furo-cho, Chikusa-ku, Nagoya 464-8602, JAPAN 
\and Theoretische Physik III, Elektronische Korrelationen und Magnetismus, Universit\"at Augsburg, Germany
\and Department of Mathematics, Imperial College, 180 Queen's Gate, London SW7 2BZ, U.K.
}
\date{Received: date / Revised version: date}
%
\abstract{
The Mott metal-insulator transition in the two-band Hubbard model in infinite dimensions is studied by using the linearized dynamical mean-field theory recently developed by Bulla and Potthoff. The phase boundary of the metal-insulator transition is obtained analytically as a function of the on-site Coulomb interaction at the $d$-orbital, the charge-transfer energy between the $d$- and $p$-orbitals and the hopping integrals between $p-d$, $d-d$ and $p-p$ orbitals. The result is in good agreement with the numerical results obtained from the exact diagonalization method. 
} 
\maketitle
%
\section{Introduction}
\label{sec:1}
 The  Mott metal-insulator transition (MIT) is a fundamental problem in the theory of strongly correlated electrons.  Recently, significant  progress has been achieved in understanding 
 this transition  by  using dynamical mean-field theory (DMFT)  \cite{Georges1}.  In the DMFT, the lattice problem is mapped into an impurity problem embedded in an effective medium by neglecting the momentum dependence of the self-energy. Various methods, such as the iterated perturbation theory \cite{Georges1}, the non-crossing approximation \cite{Pruschke}, the quantum Monte Carlo (QMC) method \cite{Jarrell}, the exact diagonalization (ED) method \cite{Caffarel} and the numerical renormalization group (NRG) method \cite{Sakai,Bulla}, enable one to solve the corresponding impurity problem. The DMFT becomes exact in the limit of infinite spatial dimensions $d=\infty$ \cite{Metzner} and is believed to be a good approximation in high dimensions.\par

The Mott MIT in the half-filled single-band Hubbard model on the $d=\infty$ Bethe lattice is found to occur as a first-order phase transition below a critical temperature $T_c \approx 0.02 W$ where $W$ is the bare bandwidth \cite{Georges1}. Below $T_c$, a coexistence of the metallic and insulating solutions was found for the same value of the on-site Coulomb interaction $U$  in the range $U_{c1}(T)<U<U_{c2}(T)$ \cite{Georges1,Rozenberg,Schlipf,Blumer,Krauth,Bulla4}. At zero temperature,  coexistence is also obtained for values of $U$ such that $U_{c1}<U<U_{c2}$. The results of the ED method for the critical values of $U$ (at $T=0$) are $U_{c1}\approx 1.2W$ and $U_{c2}\approx 1.5W$ \cite{Caffarel} (see also Section~\ref{sec:2.2}). It agrees well with the recent NRG result $U_{c1}\approx 1.25W$ and $U_{c2}\approx 1.47W$ \cite{Bulla1}. The energy of the metallic state is lower than that of the insulating state for values of $U$ in the range $U_{c1}<U<U_{c2}$. Therefore the Mott MIT occurs at $U=U_{c2}$ as a continuous transition at $T=0$. In this paper we will concentrate solely on the Mott MIT at $T=0$ and, so from this point onwards we will denote the  critical value $U_{c2}$ simply by $U_c$.\par 

The Mott MIT is observed in various $3d$ transition-metal compounds, which are classified into two types: the Mott-Hubbard type and the charge-transfer type \cite{Fujimori,Zaanen}. In the Mott-Hubbard type such as Ti and V compounds, the $d-d$ Coulomb interaction $U$ is smaller than the charge-transfer energy $\Delta$ between $d$- and anion $p$-orbitals. In this case, the energy gap of the insulator is given roughly by $U$ and a MIT  occurs at a specific value of $U$ as this interaction strength is varied. In the charge-transfer type such as Co, Ni and Cu compounds, $U$ is larger than $\Delta$. Then the energy gap is roughly given by $\Delta$ and a MIT occurs at a critical value of $\Delta$ when this energy difference is varied. \par

In the single-band Hubbard model, there is only the parameter $U/W$. The DMFT satisfactorily explains the Mott-Hubbard type MIT as a function of $U/W$. However we need at least a two-band Hubbard model with the parameters $U$ and $\Delta$ to describe the both types of MIT.\par 

In this paper we wish to study the Mott MIT with both the Mott-Hubbard type and the charge-transfer type mechanisms over the whole parameter regime. We use a  two-band Hubbard model characterized by the following parameters: the on-site Coulomb interaction $U$ at the $d$-orbital, the charge-transfer energy $\Delta$ between $d$- and $p$-orbitals and the hopping integrals $t_{pd}$, $t_{dd}$ and $t_{pp}$ between $p-d$, $d-d$ and $p-p$ orbitals, respectively. Several authors have studied the model using the DMFT approach \cite{Caffarel,Georges2,Mutou,Ono1,Ono2}. However, numerical problems make it difficult to obtain the critical values of the Mott MIT for this model, in contrast to the single-band Hubbard model. In the half-filled single-band Hubbard model on the Bethe lattice, the chemical potential is fixed to $\mu=\frac{U}{2}$ because of the particle-hole symmetry. On the other hand, for the two-band Hubbard model, the Mott MIT occurs away from particle-hole symmetry and
 so the chemical potential has to be determined explicitly to fix the electron density per unit cell to be unity. This calculation consumes a lot of CPU time.  Furthermore, the Mott MIT point is a function of several  parameters, all of which have to be calculated, making it a difficult numerical problem. \par

In this work we show that there is an alternative approach which has clear advantages as it can be handled analytically. It is based on the linearized version of DMFT, as developed by Bulla \cite{Bulla2} and applied by Bulla and Potthoff \cite{Bulla3},  where  the critical value $U_c$ of the Mott MIT  can be calculated analytically. In the linearized DMFT, the hybridization function between the impurity level and the conduction band is approximated by a single pole at the Fermi level. For the single-band Hubbard model on the Bethe lattice, the critical value is given by $U_c=1.5W$. The result is in good agreement with the numerical result from the ED and the NRG calculations of the full DMFT mentioned above. The generalization of the linearized DMFT to more complicated lattices was also discussed in ref. \cite{Bulla3}. However, the models considered there were restricted to the particle-hole symmetric case where the chemical potential is fixed to the value $\mu=\frac{U}{2}$.\par

In the present paper, we study a form of the two-band Hubbard model, which shows the Mott MIT away from the particle-hole symmetry. We generalize the linearized DMFT to the particle-hole asymmetric case and obtain an analytical expression for the critical values of the Mott MIT. A detailed account
of the calculations is given in Section~\ref{sec:2} and the analytical results are compared with the numerical results from the ED method for the several values of parameters in Section~\ref{sec:3}. A good agreement between the two approaches is found in every case. Some limiting cases in the Mott-Hubbard type and charge-transfer regimes are  discussed in Section~\ref{sec:3} and our conclusions are given in Section~\ref{sec:4}. \par

\section{Linearized dynamical mean-field theory}
\label{sec:2}
\subsection{Single-band Hubbard model}
\label{sec:2.1}

First, we consider the single-band Hubbard model, 
\begin{eqnarray} 
H= - \sum_{<i,j>,\sigma} t_{i,j} (c_{i\sigma}^{\dagger} c_{j\sigma} 
      + h.c.) + U \sum_{i} c^{\dagger}_{i\uparrow}c_{i\uparrow} 
                   c^{\dagger}_{i\downarrow}c_{i\downarrow}.  \label{HUB}
\end{eqnarray} 
In the limit of infinite dimensions, the self-energy becomes purely site-diagonal and the DMFT becomes exact. The local Green's function $G(z)$ can be given by the impurity Green's function of an effective single impurity Anderson model, 
\begin{eqnarray} 
H_{\rm And}&=& \varepsilon_f \sum_{\sigma} f^{\dagger}_\sigma f_\sigma + U f^{\dagger}_{\uparrow}f_{\uparrow} f^{\dagger}_{\downarrow}f_{\downarrow} \nonumber \\ 
&+&\sum_{k,\sigma} \varepsilon_k c^{\dagger}_{k\sigma} c_{k\sigma} +\sum_{k,\sigma}V_k(f^{\dagger}_{\sigma}c_{k\sigma}+c^{\dagger}_{k\sigma}f_{\sigma}), 
\label{AND}
\end{eqnarray} 
where $\varepsilon_f$ is the impurity level and $\varepsilon_k$ are energies of conduction electrons hybridized with the impurity by $V_k$. 
In the model eq. (\ref{AND}), the non-interacting impurity Green's function, 
\begin{eqnarray} 
{\cal G}_0(z) = (z-\varepsilon_f- \Delta(z))^{-1}, \label{G0}
\end{eqnarray} 
with the hybridization function, 
\begin{eqnarray} 
\Delta(z)=\sum_k \frac{V_k^2}{z-\varepsilon_k}, \label{DELTA}
\end{eqnarray} 
includes effects of the interaction at all the sites except the impurity site and is determined self-consistently so as to satisfy the self-consistency equation. 

For simplicity, the calculations in this paper are restricted to the Bethe lattice with the connectivity $q$ and the hopping $t_{i,j}=\frac{t}{\sqrt{q}}$ \cite{Bethe}. In the limit $q=\infty$, the self-consistent equation is given by 
\begin{equation} 
{\cal G}_0(z)^{-1} = z+\mu-t^2 G(z), \label{SCE}
\end{equation} 
where $\mu$ is the chemical potential for the original lattice model. 
In the non-interacting case, the local Green's function is obtained from eq.(\ref{SCE}) with ${\cal G}_0(z)=G(z)$. It yields a semicircular density of states: 
$
D(\varepsilon)=\frac{1}{\pi t}\sqrt{1-(\frac{\varepsilon+\mu}{2t})^2} 
$
for $|\varepsilon+\mu|<2t$. 
Because of the particle-hole symmetry at half-filling, the chemical potential and the impurity level are set to $\mu = \frac{U}{2}$ and $\varepsilon_f = -\frac{U}{2}$, respectively. Then, the self-consistency equation (\ref{SCE}) is simply written by 
\begin{eqnarray} 
\Delta(z)=t^2 G(z). \label{SCE2}
\end{eqnarray}

When the system approaches the MIT from the metallic side at $T=0$, the central quasiparticle peak is found to appear to be isolated from the upper and the lower Hubbard bands \cite{Georges1,Bulla1}. The width of the quasiparticle peak vanishes in the limit $U\to U_c$. In this limit, the effect of the Hubbard bands on the quasiparticle peak becomes rather small \cite{Bulla3}. Therefore, Bulla and Potthoff \cite{Bulla3} used an approximate form for the hybridization function where the contribution from the Hubbard bands are completely removed and that the quasiparticle peak is replaced by a single pole at the Fermi level \cite{Fermi}, 
\begin{eqnarray} 
\Delta(z)=\frac{\Delta_0}{z}, \label{DELTA0}
\end{eqnarray} 
with the small weight $\Delta_0$ which will be determined self-consistently. 
This model with eq.(\ref{DELTA0}) corresponds to the two-site Anderson model \cite{Hewson}, 
\begin{eqnarray} 
H_{\rm 2-site}&=& \varepsilon_f \sum_{\sigma} f^{\dagger}_\sigma f_\sigma +U f^{\dagger}_{\uparrow}f_{\uparrow}f^{\dagger}_{\downarrow}f_{\downarrow} \nonumber \\ 
&+&\varepsilon_c \sum_{\sigma} c^{\dagger}_{\sigma} c_{\sigma} + V \sum_{\sigma}(f^{\dagger}_{\sigma}c_{\sigma}+c^{\dagger}_{\sigma}f_{\sigma}), 
\label{2SITE}
\end{eqnarray} 
with $V=\sqrt{\Delta_0}$, $\varepsilon_c=0$ and $\varepsilon_f=-\frac{U}{2}$. The model is solved analytically to obtain the impurity Green's function which has four poles: two poles at $\omega \approx \pm\frac{U}{2}$ and two poles near the Fermi level $\omega\approx 0$ when $U$ is large. These latter poles are precursors of the Kondo resonance in the Anderson model with complete conduction band.

When $U \to U_c$, the hybridization becomes $V=\sqrt{\Delta_0} \to 0$. 
In this limit, the impurity, and therefore, the local Green's function is given by $G(z) = w/z$ near the Fermi level with the residue \cite{Hewson} (see also Appendix \ref{Apex:1}),
\begin{eqnarray} 
w = 36 \frac{V^2}{U^2} = 36 \frac{\Delta_0}{U^2}, \label{W}
\end{eqnarray}
up to the second order in $V$.
From eqs. (\ref{SCE2}) and (\ref{W}), we obtain a new hybridization function which has a pole at $z=0$ with the weight, 
\begin{eqnarray} 
\Delta_0' = 36 \frac{t^2}{U^2} \Delta_0. \label{W2}
\end{eqnarray}
When we solve the self-consistency equation (\ref{SCE2}) by iteration, $\Delta_0$ for the $(N+1)$th iteration step is expressed in terms of $\Delta_0$ for the $N$th step through eq.(\ref{W2}). Therefore the critical value for the MIT is given by 
\begin{eqnarray} 
U_c = 6t=1.5W, \label{UC}
\end{eqnarray}
with the bare bandwidth $W=4t$. 
For $U<U_c$, the weight $\Delta_0$ increases exponentially with iteration number and, then, the single pole approximation for $\Delta(z)$  breaks down. For $U>U_c$, $\Delta_0$ decreases exponentially to obtain the self-consistent value $\Delta_0 = 0$ corresponding to the insulating solution.

\subsection{Comparison with numerical methods}
\label{sec:2.2}

\begin{figure*}
\begin{center}
\resizebox{0.4\textwidth}{!}{
\includegraphics{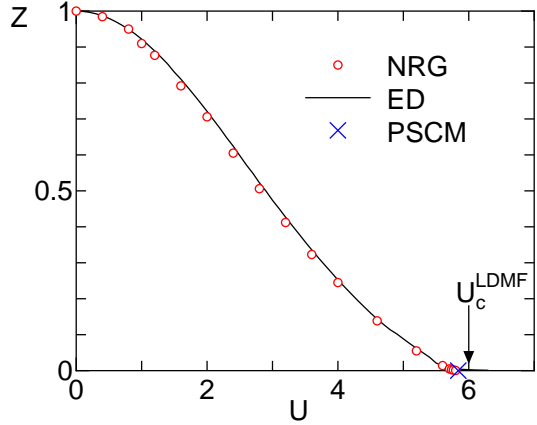}
}
\end{center}
\caption{
The quasiparticle weight $Z$ in the single-band Hubbard model on the Bethe lattice as a function of $U$: a comparison between the NRG result \cite{Bulla1} and the ED result (the system size $n_s=8$). The critical value from the linearized DMFT, $U_c^{\rm LDMF}$, and that from the PSCM \cite{Georges1} are also shown. We set $t=1$.
}
\label{fig_z}       
\end{figure*}

Here we estimate the reliability of the linearized DMFT by comparing the analytical result of the critical value eq.(\ref{UC}) with the available numerical results. 
The effective single impurity Anderson model eq.(\ref{AND}) is approximately solved by using the exact diagonalization of a cluster model with finite system size $n_s$ \cite{Caffarel} (the ED method). 
The Wilson's numerical renormalization group method is also used to solve the model eq.(\ref{AND}) in the thermodynamic limit \cite{Sakai,Bulla} (the NRG method). 

Fig.~\ref{fig_z} shows the $U$ dependence of the quasiparticle weight defined by 
$ 
Z=(1-\frac{d\Sigma (z)}{d z}|_{z=0} )^{-1},
$ 
with the local self-energy
$ 
\Sigma (z)={\cal G}_0(z)^{-1} - G(z)^{-1}, 
$ 
calculated from the NRG method \cite{Bulla1} and the ED method. 
The NRG and the ED results agree very well over the whole range of $U$-values. 

When we approach the MIT point from the metallic side, $Z$ continuously becomes zero at a critical value $U_c$ as shown in Fig.~\ref{fig_z}. 
Recently, Bulla \cite{Bulla1} obtained the precise result of the critical value $U_c=5.88t$ by using the NRG method. In the ED method, an extrapolation of the systems with up to $n_s=11$ yields the $n_s \to \infty$ extrapolated value $U_{c}=5.87t$ \cite{Ono3}. 
The result from the linearized DMFT, eq.(\ref{UC}), is in very good agreement with the NRG and the ED results. It also agrees well with the value of $U_c=5.84t$ from the projective self-consistent method (PSCM) \cite{Georges1} and with the value of $U_c=6.04t$ obtained in the NRG calculations of Shimizu and Sakai \cite{Shimizu}. The iterated perturbation method, where the effective single impurity Anderson model eq.(\ref{AND}) is solved within the second order perturbation with respect to $U$, gives a larger critical value $U_c=6.6t$ as compared to the other non-perturbative approaches. The random dispersion approximation (RDA) \cite{Noack} predicts a considerably lower critical value $U_c=4.0t$. The origin of this discrepancy is presently not clear.

\subsection{Two-band Hubbard model}
\label{sec:2.3}

Next, we consider the two-band Hubbard model on a Bethe lattice with connectivity $q$, 
\begin{eqnarray} 
H &=& \frac{t_{pd}}{\sqrt{q}} \sum_{<i,j>,\sigma}
        ( d_{i\sigma}^{\dagger} p_{j\sigma} + h.c.)
      + U \sum_{i} d^{\dagger}_{i\uparrow}d_{i\uparrow} 
                   d^{\dagger}_{i\downarrow}d_{i\downarrow}  
        \nonumber \\
  &+&  \frac{t_{dd}}{q} \sum_{<i,i'>,\sigma} 
       (d_{i\sigma}^{\dagger}d_{i'\sigma} + h.c.)
     + \varepsilon_d \sum_{i,\sigma} d_{i\sigma}^{\dagger} d_{i\sigma}
 \nonumber \\
  &+&  \frac{t_{pp}}{q} \sum_{<j,j'>,\sigma} 
       (p_{j\sigma}^{\dagger}p_{j'\sigma} + h.c.)
     + \varepsilon_p \sum_{j,\sigma} p_{j\sigma}^{\dagger} p_{j\sigma},
        \label{DP}
\end{eqnarray} 
where $d^{\dagger}_{i\sigma}$ and $p_{j\sigma}^{\dagger}$ are creation operators for an electron with spin $\sigma$ in the $d$-orbital at site $i$ and in the $p$-orbital at site $j$, respectively. 
$t_{pd}$, $t_{dd}$ and $t_{pp}$ are the hopping integrals between the nearest neighbour $p-d$, $p-p$ and $d-d$ orbitals, respectively. The charge-transfer energy $\Delta$ is defined by $\Delta \equiv \varepsilon_{p}-\varepsilon_{d}>0$ \cite{Hole}. In eq.(\ref{DP}), we assume that $p$- and $d$-orbitals are on different sub-lattices of a bipartite lattice, more explicitly, a Bethe lattice with connectivity $q$. 
In the limit $q = \infty$, the self-consistency equations for the local Green's functions are given by \cite{Georges1,Georges2}
\begin{eqnarray} 
{\cal G}_0(z)^{-1}&=& z +\mu -\varepsilon_d -t_{pd}^2 G_p(z) -t_{dd}^2 G_d(z),
 \label{SCE3A} \\
G_p(z)^{-1}&=& z + \mu - \varepsilon_p - t_{pd}^2 G_d(z) -t_{pp}^2 G_p(z), 
 \label{SCE3B}
\end{eqnarray} 
where $G_p(z)$ is the local Green's function for the $p$-electron and $G_d(z)$ is that for the $d$-electron which can be given by the impurity Green's function of an effective single impurity Anderson model eq.(\ref{AND}).

\begin{figure*}
\begin{center}
\resizebox{0.27\textwidth}{!}{
\includegraphics{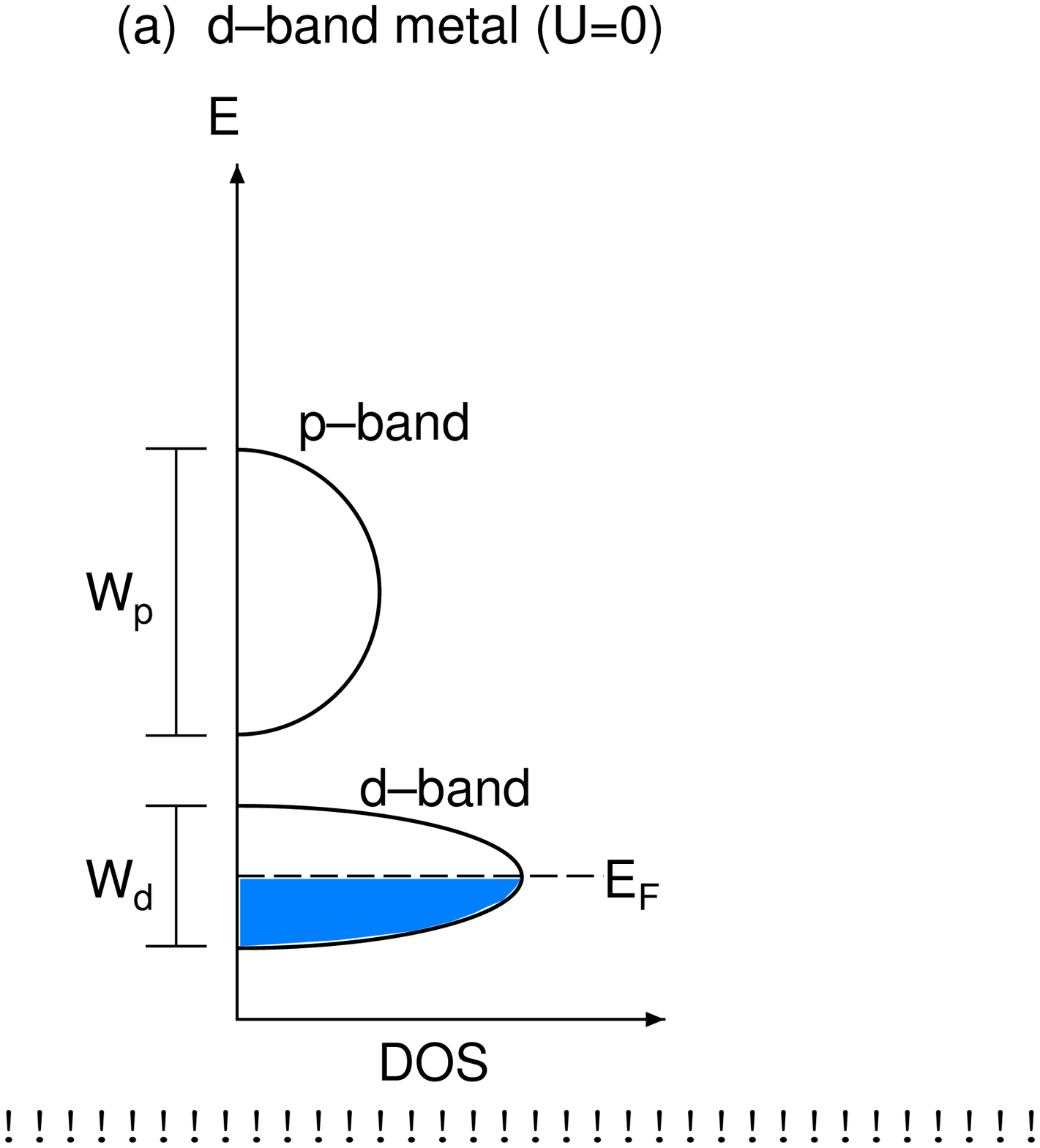}
}
\resizebox{0.27\textwidth}{!}{
\includegraphics{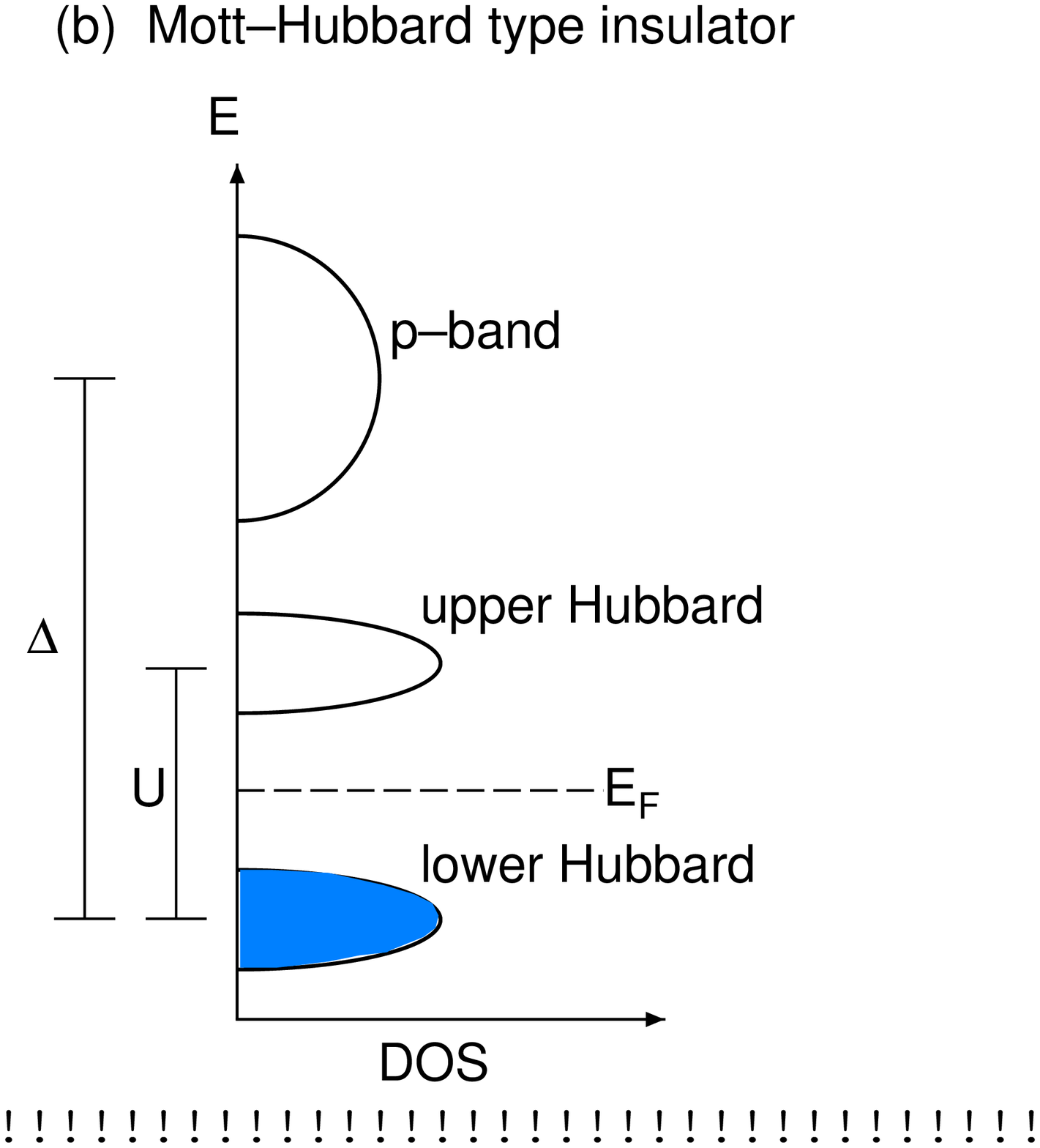}
}
\resizebox{0.27\textwidth}{!}{
\includegraphics{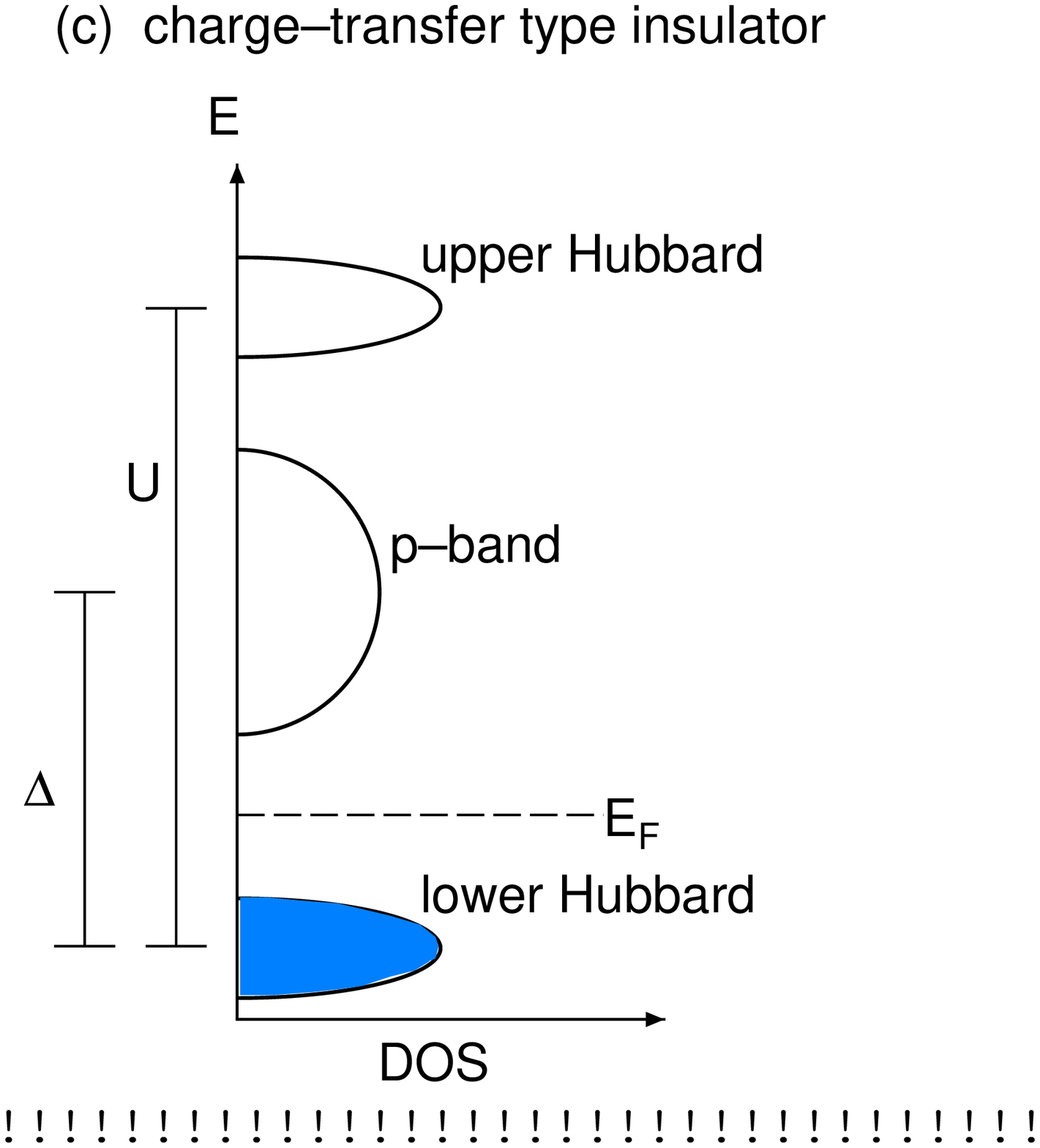}
}
\end{center}
\caption{
Schematic figures of the density of states in the two-band Hubbard model: (a) the $d$-band metal with $U=0$, (b) the Mott-Hubbard type insulator with $U<\Delta$ and (c) the charge-transfer type insulator with $U>\Delta$ \cite{Hole}. $E_F$ is the Fermi level.
}
\label{fig_dos}       
\end{figure*}

The non-interacting local Green's functions are easily obtained by solving the self-consistency equations (\ref{SCE3A}, \ref{SCE3B}) with ${\cal G}_0(z)=G_d(z)$. They yield the densities of states which consist of two bands: a $d$-band with the bandwidth $W_d\approx 4t_{dd}$ around the $d$-level and a $p$-band with the bandwidth $W_p\approx 4t_{pp}$ around the $p$-level when $t_{pd}$ is small. When the electron density per unit cell $n=1$, the $d$-band is half-filled and the system is metallic as shown in Fig.~\ref{fig_dos}(a). 

In the presence of $U$, the system is found to show the Mott MIT \cite{Georges2,Mutou,Ono1,Ono2}. 
When $U<\Delta$ (Mott-Hubbard type), the energy gap of the insulating state is approximately given by $U$ (Fig.~\ref{fig_dos}(b)). On the other hand, when $U>\Delta$ (charge-transfer type), it is approximately given by $\Delta$ (Fig.~\ref{fig_dos}(c)). 
When the system approaches the MIT from the metallic side, the central quasiparticle peak is found to be largely isolated from the upper and lower Hubbard bands in the Mott-Hubbard type as observed in the single-band Hubbard model. It is also found to be largely isolated from the $p$-band and the lower Hubbard band in the charge-transfer type \cite{Mutou,Ono1}. The width of the quasiparticle peak vanishes in the limit of the MIT point in both types.

Now, we study the MIT of this system by using the linearized DMFT. We assume that in the limit of the MIT point the effect of both the Hubbard bands and the $p$-band on the quasiparticle peak is negligible. Then we use an approximate form for the hybridization function eq.(\ref{DELTA0}) as used in the single-band Hubbard model. In this case, the effective single impurity Anderson model eq.(\ref{AND}) with  eq.(\ref{DELTA0}) corresponds to the two-site Anderson model eq.(\ref{2SITE}) with $V=\sqrt{\Delta_0}$, $\varepsilon_c=0$ and 
$\varepsilon_f=\bar\varepsilon_d\equiv\varepsilon_d-\mu$. 
The local $d$-Green's function in this model is obtained by (see Appendix \ref{Apex:1})
\begin{eqnarray} 
G_d(z)=\frac{w_d}{z}+\frac{w_1}{z-\varepsilon_1}+\frac{w_4}{z-\varepsilon_4}, 
   \label{GD}
\end{eqnarray} 
where the residue $w_d$ is (up to the second order in $V$) 
\begin{eqnarray} 
w_d = V^2\left( \frac{5}{2\bar\varepsilon_d^2} 
             + \frac{5}{2(\bar\varepsilon_d+U)^2}
             - \frac{4}{\bar\varepsilon_d(\bar\varepsilon_d+U)} \right), 
         \label{WD}
\end{eqnarray} 
and $w_1$, $w_4$, $\varepsilon_1$ and $\varepsilon_4$ are given in Appendix \ref{Apex:1}. 

To calculate the local $p$-Green's function, we assume an approximate form \cite{Approx} 
\begin{eqnarray} 
G_p(z)=\frac{w_p}{z} + \frac{1-w_p}{z-\bar\varepsilon_p}, 
         \label{GP}
\end{eqnarray} 
with $\bar\varepsilon_p\equiv\varepsilon_p-\mu$. By using eq.(\ref{GP}) in eq.(\ref{SCE3B}) together with eq.(\ref{GD}), we obtain $w_p$ up to the second order in $V$: 
\begin{eqnarray} 
w_p = \frac{t_{pd}^2 w_d}{
      \left(\bar\varepsilon_p -\frac{t_{pd}^2}{2\bar\varepsilon_d}
   -\frac{t_{pd}^2}{2(\bar\varepsilon_d+U)}-\frac{t_{pp}^2}{\bar\varepsilon_p}
          \right)^2 -t_{pp}^2}. 
      \label{WP}
\end{eqnarray} 

Substituting eqs.(\ref{GD}, \ref{GP}) into eq.(\ref{SCE3A}) yields a new hybridization function which has a pole at $z=0$ with the weight 
$\Delta_0'= F\Delta_0$ with 
\begin{eqnarray} 
F(t_{pd},t_{pp},t_{dd},U,\bar\varepsilon_d,\bar\varepsilon_p) 
          \hspace{40mm}  \nonumber \\
   = \left( \frac{5}{2\bar\varepsilon_d^2} 
             + \frac{5}{2(\bar\varepsilon_d+U)^2}
             - \frac{4}{\bar\varepsilon_d(\bar\varepsilon_d+U)} \right)
          \hspace{10mm}  \nonumber \\
   \times \left(  \frac{t_{pd}^4}{ 
         \left(\bar\varepsilon_p -\frac{t_{pd}^2}{2\bar\varepsilon_d}
    -\frac{t_{pd}^2}{2(\bar\varepsilon_d+U)}-\frac{t_{pp}^2}{\bar\varepsilon_p}
          \right)^2 -t_{pp}^2} +t_{dd}^2 
            \right). 
\label{F}
\end{eqnarray} 
Following the same argument discussed in Section \ref{sec:2.1}, we have an equation to determine the MIT point within the linearized DMFT: 
\begin{eqnarray} 
F(t_{pd},t_{pp},t_{dd},U,\bar\varepsilon_d,\bar\varepsilon_p)=1. 
\label{F1}
\end{eqnarray} 

In eq.(\ref{F1}), $F$ includes the chemical potential $\mu$ (through $\bar\varepsilon_d$ and $\bar\varepsilon_p$). 
In the two-band Hubbard model, the MIT occurs away from the particle-hole symmetry as shown in Fig.~\ref{fig_dos}. Then we have to determine the chemical potential explicitly to obtain the critical values of the MIT. In general, the chemical potential is determined so as to fix the electron density. 
In the linearized DMFT, we focus on the low-energy part of the Green's function and determine it self-consistently. On the other hand, the high-energy part, whose details are neglected, is not determined self-consistently. Then we fail to obtain the precise expression of the electron density calculated from the Green's function within the linearized DMFT. However, in the next paragraph we show that we can use an alternative  condition to determine $\mu$, based on the fact that at the MIT point $\Delta_0'$ has a minimum value as a function of $\mu$. 
This condition gives
\begin{eqnarray} 
\frac{\partial}{\partial \mu}
    F(t_{pd},t_{pp},t_{dd},U,\bar\varepsilon_d,\bar\varepsilon_p) = 0. 
\label{F2}
\end{eqnarray} 
Combined use of eq.(\ref{F1}) and eq.(\ref{F2}) with eq.(\ref{F}) enables us to obtain an analytic expression for the phase boundary separating the metallic and insulating regimes as a function of the parameters: $t_{pd}$, $t_{pp}$, $t_{dd}$, $U$ and $\Delta\equiv\varepsilon_{p}-\varepsilon_{d}$. 

In the metallic regime, the chemical potential $\mu(n)$ is continuous at $n=1$ as a function of $n$. On the other hand, in the insulating regime, $\mu(n)$ has a jump at $n=1$. When we approach the MIT phase boundary from the metallic side, $\mu(n)$ is still continuous even in the limit of the MIT point. Correspondingly, there are three cases in the $\mu$ dependence of $\Delta_0'(\mu)$ as below. 
(1) In the metallic regime, 
$\Delta_0'(\mu)>\Delta_0$ for all $\mu$ resulting in the metallic solution for all $n$. 
(2) In the insulating regime, 
$\Delta_0'(\mu)<\Delta_0$ for $\mu_- < \mu < \mu_+$, 
while, $\Delta_0'(\mu)>\Delta_0$ for $\mu < \mu_-$ or $\mu > \mu_+$. 
Then the system is a Mott insulator for $\mu_- < \mu < \mu_+$, and $\mu$ shows a jump from $\mu_-$ to $\mu_+$ at $n=1$. 
(3) On the phase boundary of the MIT, 
$\Delta_0'(\mu)=\Delta_0$ for $\mu = \mu(n=1)$, 
while, $\Delta_0'(\mu)>\Delta_0$ for $\mu \ne \mu(n=1)$. 
Then $\Delta_0'(\mu)$ has a minimum at $\mu = \mu(n=1)$. 
Therefore the equation (\ref{F2}) is the unique condition to determine the chemical potential on the MIT phase boundary within the linearized DMFT. 

In the single-band Hubbard model on the Bethe lattice, the condition to minimize $\Delta_0'$ is written by $\frac{\partial w_d}{\partial\mu}=0$ where $w_d$ is defined by eq.(\ref{WD}) with $\varepsilon_d=0$ ($\bar\varepsilon_d=-\mu$) without assuming the particle-hole symmetry. It yields the exact value of the chemical potential, $\mu=\frac{U}{2}$, as expected. In the two-band Hubbard model, the chemical potential thus obtained agrees well with that from the ED method where the high-energy part is also determined self-consistently. It will be shown in Section~\ref{sec:3}.

\section{Discussion}
\label{sec:3}

\subsection{Phase diagram}
\label{sec:3.1}

\begin{figure*}
\begin{center}
\resizebox{0.4\textwidth}{!}{
\includegraphics{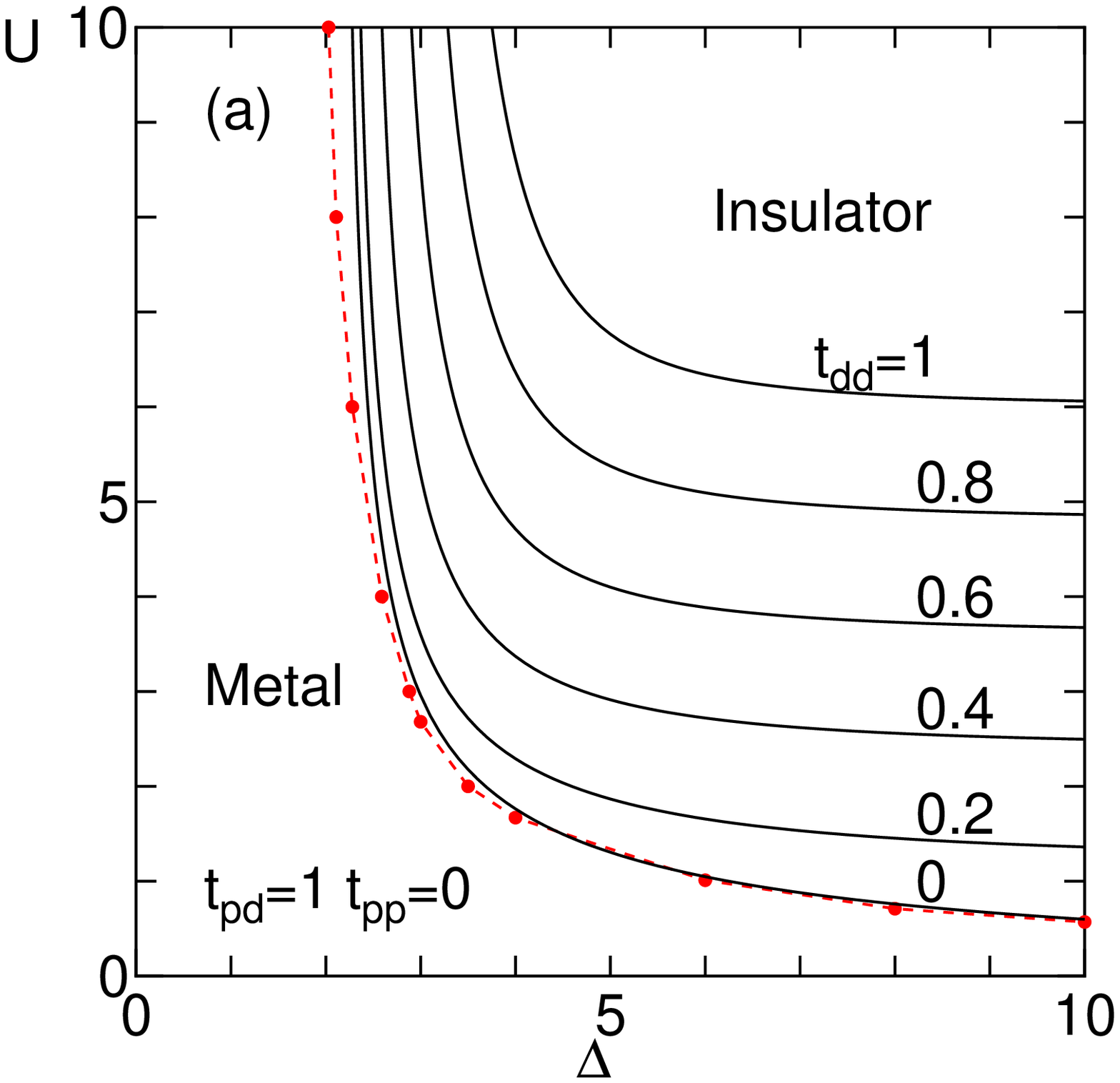}
}
\resizebox{0.4\textwidth}{!}{
\includegraphics{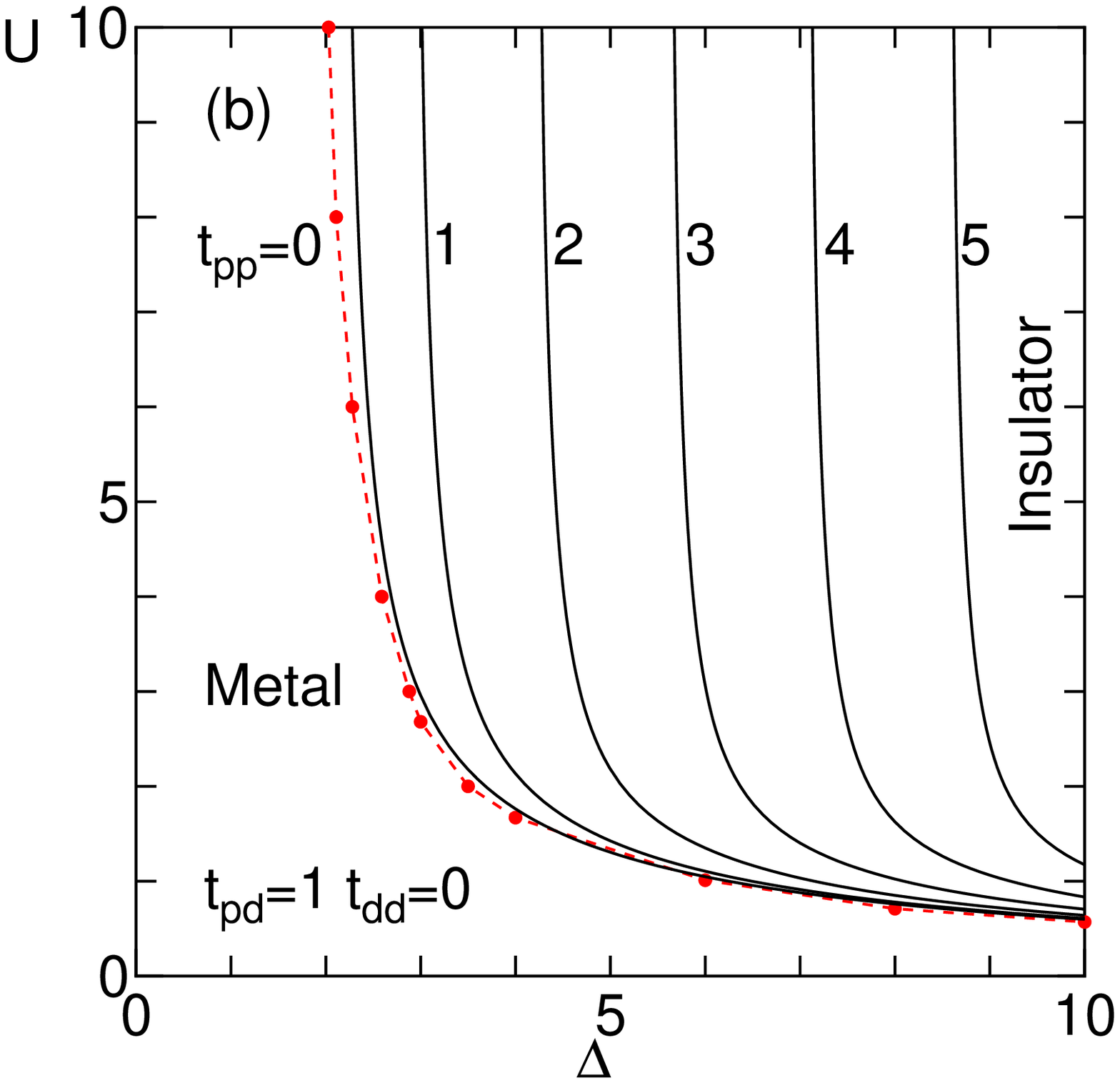}
}
\resizebox{0.4\textwidth}{!}{
\includegraphics{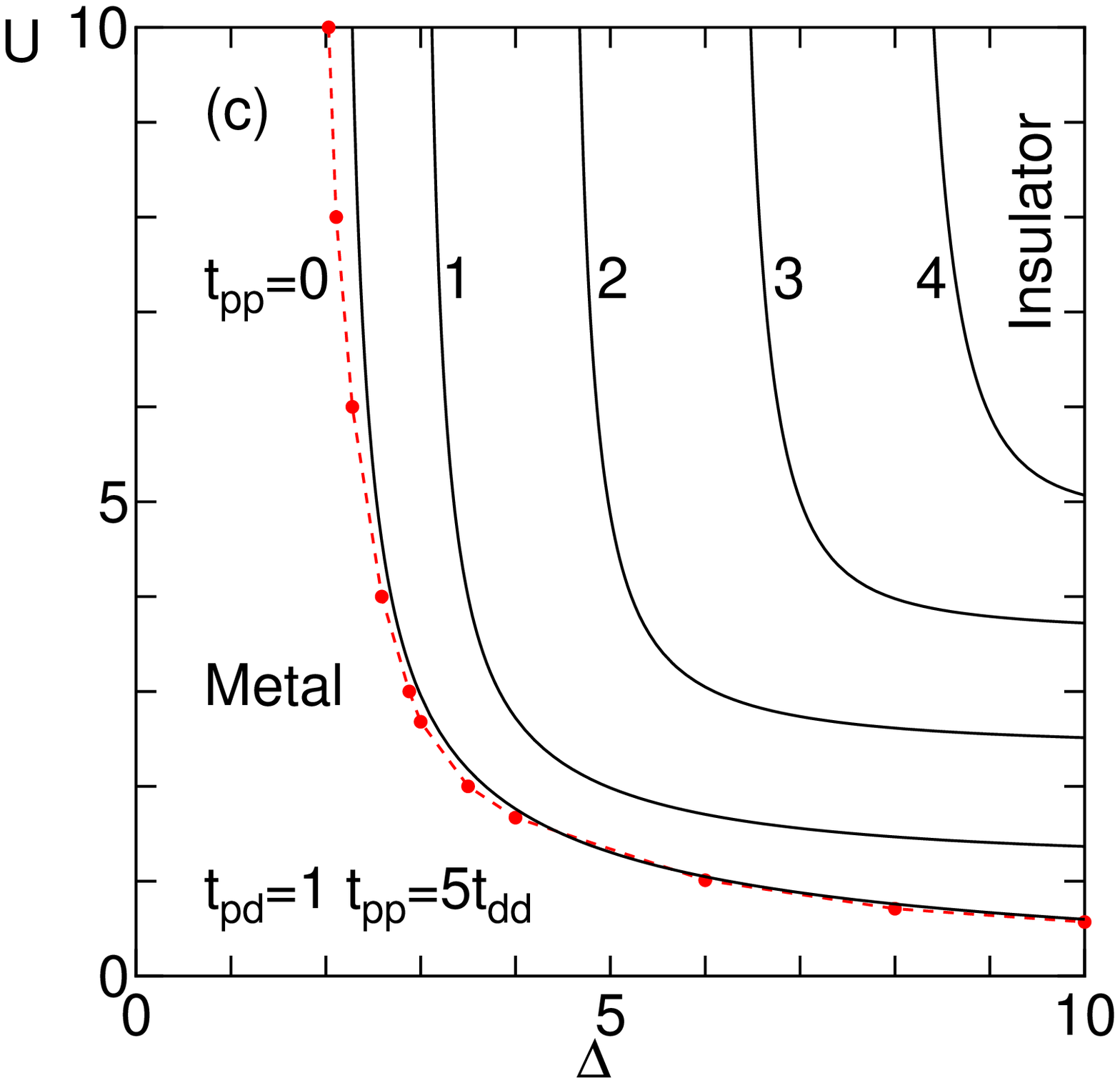}
}
\end{center}
\caption{
Phase diagrams of the two-band Hubbard model at $T=0$ and $n=1$. Solid lines are phase boundaries separating the metallic and insulating regimes obtained from the linearized DMFT as functions of $\Delta$ and $U$ for several values of $t_{dd}$ with $t_{pp}=0$ (a), for several values of $t_{pp}$ with $t_{dd}=0$ (b) and for several values of $t_{pp}=5t_{dd}$ (c). Closed circles are the critical values for $t_{pp}=t_{dd}=0$ calculated from the exact diagonalization method \cite{Ono2}. We set $t_{pd}=1$ in all figures. 
}
\label{fig_p}       
\end{figure*}

\begin{figure*}
\begin{center}
\resizebox{0.4\textwidth}{!}{
\includegraphics{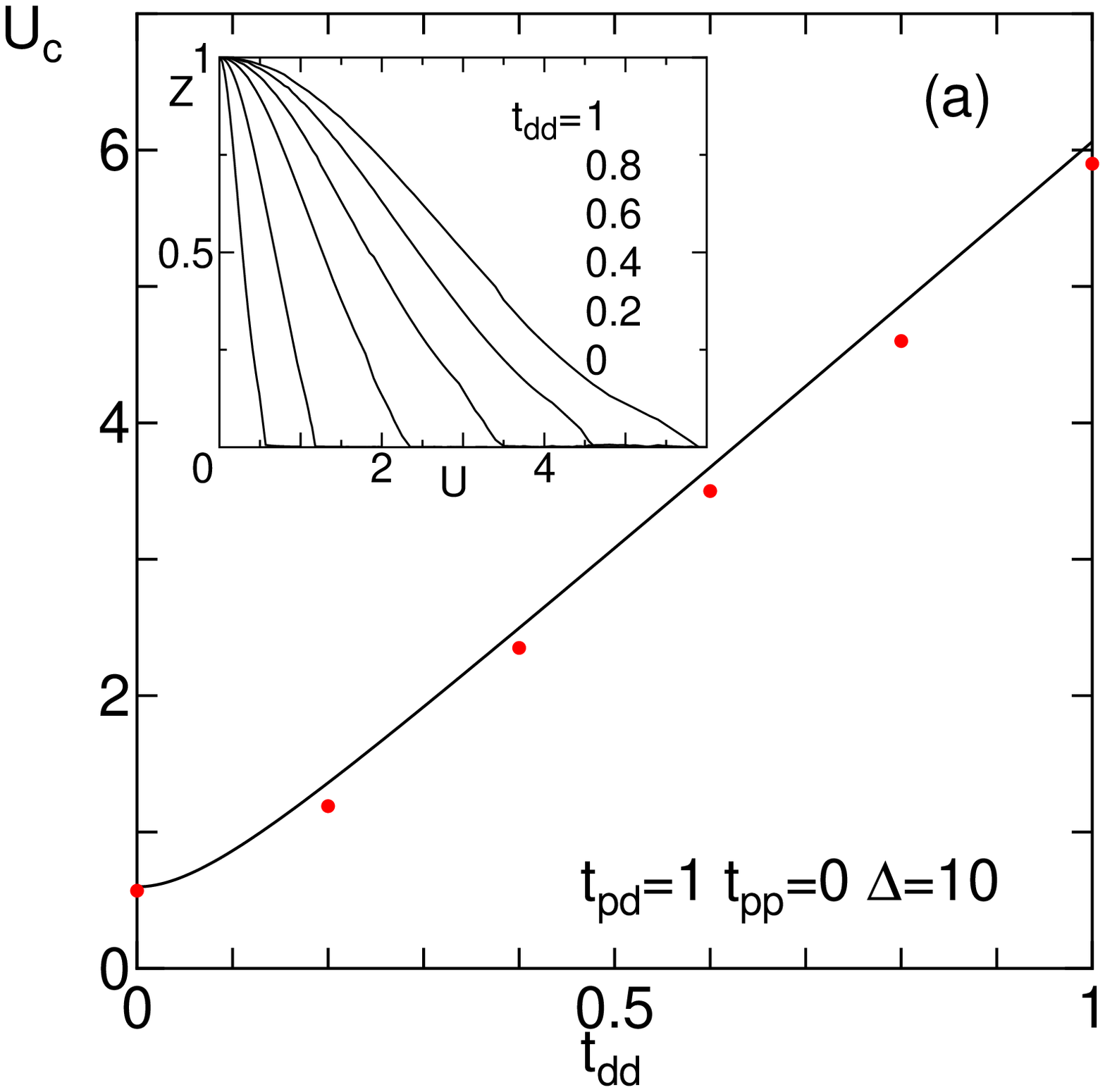}
}
\resizebox{0.4\textwidth}{!}{
\includegraphics{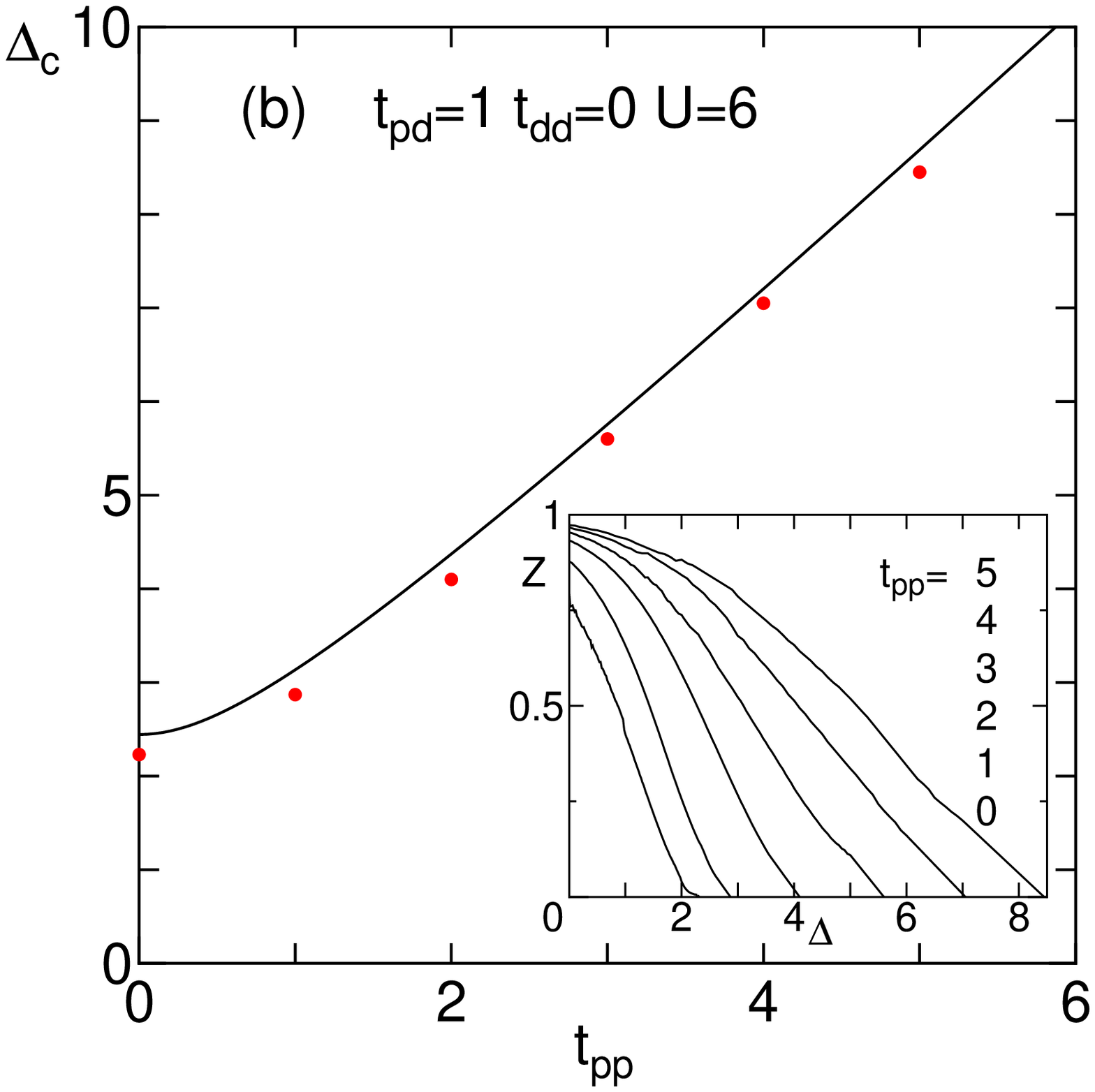}
}
\end{center}
\caption{
The critical value $U_c$ as a function of $t_{dd}$ for $t_{pp}=0$ and $\Delta=10$ (a), and the critical value $\Delta_c$ as a function of $t_{pp}$ for $t_{dd}=0$ and $U=6$ (b), obtained from the linearized DMFT (solid line) and from the ED method (closed circle). Insets show the quasiparticle weight $Z$ from the ED method (the system size $n_s=8$) as a function of $U$ for several $t_{dd}$ with $t_{pp}=0$ and $\Delta=10$ (a), and as a function of $\Delta$ for several $t_{pp}$ with $t_{dd}=0$ and $U=6$ (b). We set $t_{pd}=1$ in all figures. 
}
\label{fig_uc}       
\end{figure*}

From the coupled equations (\ref{F1},\ref{F2}) with eq.(\ref{F}), we can easily obtain the phase boundary of the MIT as a function of $t_{pd}$, $t_{pp}$, $t_{dd}$, $U$ and $\Delta$ within the linearized DMFT. Figs.~\ref{fig_p}(a)-(c) show the phase diagrams of the MIT on the $\Delta-U$ plane for several values of $t_{pp}$ and $t_{dd}$ with $t_{pd}=1$ together with the result obtained from the ED method for $t_{pp}=t_{dd}=0$ \cite{Ono2}. The result from the linearized DMFT is in good agreement with the ED result for all values of $\Delta$ and $U$ in the case with $t_{pp}=t_{dd}=0$.

The MIT is observed when $U$ is varied for $U<\Delta$ (Mott-Hubbard type), while it is observed when $\Delta$ is varied for $U>\Delta$ (charge-transfer type) as seen in Figs.~\ref{fig_p}(a)-(c). 
The phase boundary smoothly connects the Mott-Hubbard type and the charge-transfer type transitions for all values of $t_{dd}$ and $t_{pp}$. 
As $t_{dd}$ and/or $t_{pp}$ increase, the metallic region monotonically increases. However, the effect of $t_{pp}$ on the critical value of the Mott-Hubbard type transition is rather small as seen in Fig.~\ref{fig_p}(b). This will be discussed in Section~\ref{sec:3.3}.

In the presence of $t_{pp}$ and $t_{dd}$, we also calculated the quasiparticle weight $Z$ by using the ED method.  When we approach the MIT point from the metallic side, $Z$ continuously becomes zero as shown in the insets of Figs.~\ref{fig_uc}(a) and (b). The critical value $U_c$ ($\Delta_c$) thus obtained is plotted in Fig.~\ref{fig_uc}(a) (Fig.~\ref{fig_uc}(b)) as a function of $t_{dd}$ ($t_{pp}$) for a fixed value of $\Delta$ ($U$) together with that obtained from the linearized DMFT. The agreement between the two methods is good even in the case with finite $t_{pp}$ and $t_{dd}$. Thus we conclude that the linearized DMFT gives a reliable estimate for the phase boundary of the Mott MIT in the two-band Hubbard model over the whole parameter regime. 
We note that the chemical potential at the MIT point from the linearized DMFT also agrees well with that from the ED method. This confirms that eq.(\ref{F2}) is a reliable condition to determine the chemical potential at the MIT point.

By eliminating the chemical potential $\mu$ directly from the coupled equations (\ref{F1},\ref{F2}) with eq.(\ref{F}), an analytic expression for the phase boundary of the MIT is obtained within the linearized DMFT, although it is rather complicated. We can however get some simple analytical expression in limiting cases as below.

\subsection{Mott-Hubbard regime ($U<\Delta$)}
\label{sec:3.2}

For the case with $U<\Delta$, the MIT is observed when $U$ is varied as seen in Figs.~\ref{fig_p}(a)-(c). In the limit $\Delta \to \infty$, eqs.(\ref{F1},\ref{F2}) with eq.(\ref{F}) yield the chemical potential $\mu \approx \frac{U_c}{2}$ and the critical value $U_c$: 
\begin{eqnarray} 
U_c \approx 6\sqrt{\frac{t_{pd}^4}{\Delta^2}+t_{dd}^2}.
\label{UC2}
\end{eqnarray} 
When $t_{pd}=0$, eq.(\ref{UC2}) results in $U_c\approx 6t_{dd}=1.5W_d$ with the bare $d$-bandwidth $W_d=4t_{dd}$ which is equivalent to eq.(\ref{UC}) in the case with single-band Hubbard model as expected. When $t_{dd}=0$, eq.(\ref{UC2}) yields $U_c \approx 1.5W_{d}$ with the bare $d$-bandwidth of the $d$-$p$ hybridized band given by 
$
W_{d}=\frac12 [(\Delta^2+16t_{pd}^2)^{\frac12}-\Delta] \approx \frac{4t_{pd}^2}{\Delta}. 
$
In general, $U_c$ is approximately given by 1.5 times the bare $d$-bandwidth in the Mott-Hubbard regime. 
We note that, in the limit $\Delta\to \infty$, $U_c$ is independent of $t_{pp}$ as shown in eq.(\ref{UC2}) and also shown in Fig.~\ref{fig_p}(b). This enables us to describe the MIT in the Mott-Hubbard type by using the single-band Hubbard model.

\subsection{Charge-transfer regime ($U>\Delta$)}
\label{sec:3.3}

For the case with $U > \Delta$, the MIT is observed when $\Delta$ is varied as seen in Figs.~\ref{fig_p}(a)-(c). In the limit $U \to \infty$, eqs.(\ref{F1},\ref{F2}) with eq.(\ref{F}) yield the critical value $\Delta_c$: 
\begin{eqnarray} 
\Delta_c \approx \left\{
\begin{array}{ll}
2\left(\sqrt{\frac52} -\frac12\right)^\frac12 t_{pd} \ \ \ &   t_{pd}\gg t_{pp}, t_{dd}, \\
\frac{1+\sqrt{5}}{2} t_{pp} & t_{pp}\gg t_{pd}, t_{dd}, \\
\sqrt{\frac52} t_{dd} + \frac{2t_{dd}^2}{U} & t_{dd}\gg t_{pd}, t_{pp}. 
\end{array}
\right. \label{DC}
\end{eqnarray}
Correspondingly, the chemical potential is given by 
$\mu \approx \frac{\varepsilon_p+\varepsilon_d}{2}$
for $t_{pd}\gg t_{pp}, t_{dd}$, 
$\mu \approx \varepsilon_d$
for $t_{pp}\gg t_{pd}, t_{dd}$ and 
$\mu \approx \varepsilon_p$
for $t_{dd}\gg t_{pd}, t_{pp}$. 
We note that for the small value of $\Delta <\Delta_c$ the system is metallic even in the limit $U\to \infty$ as mentioned in ref. \cite{Zaanen}. When $U \gg \Delta \gg t_{pd}, t_{pp}, t_{dd}$, $d$-orbitals are almost singly occupied and $p$-orbitals are nearly empty. The electron transfer from a $d$-orbital to a $p$-orbital costs the charge-transfer energy $\Delta$ while it gains the kinetic energy: 
$K\sim \frac{t_{pd}^2}{\Delta}$
for $t_{pd}\gg t_{pp}, t_{dd}$, 
$K\sim t_{pp}$
for $t_{pp}\gg t_{pd}, t_{dd}$ and 
$K\sim t_{dd}+O(\frac{t_{dd}^2}{U})$
for $t_{dd}\gg t_{pd}, t_{pp}$. 
Equation (\ref{DC}) means an energy balance between the charge-transfer energy and the kinetic energy \cite{Energy}. 

For general values of $t_{pd}, t_{pp}$ and $t_{dd}$, the explicit expression for $\Delta_c$ is still complicated even in the limit $U \to \infty$. However the critical value $\Delta_c$ is roughly given by the energy balance mentioned above in the charge-transfer regime. We note that $\Delta_c$ is positive for all values of $U$, $t_{pd}, t_{pp}$ and $t_{dd}$ in contrast to the negative-$\Delta_c$ predicted by the local impurity approximation \cite{Mizokawa} where the impurity Anderson model is solved (not self-consistently) to determine the critical value of the MIT. This descrepancy will be discussed in Section~\ref{sec:4}.

\section{Conclusions}
\label{sec:4}

Within the linearized DMFT, we have obtained a good description of the phase diagram of the Mott MIT analytically in the two-band Hubbard model over the whole parameter regime including the Mott-Hubbard regime, charge-transfer regime and the intermediate regime. The analytical result agrees well with the numerical result obtained from the ED method. Although the ED method is an approximate calculation, it is in very good agreement with the recent NRG method in the single-band Hubbard model and is expected to be reliable for the two-band Hubbard model.

We have used the two-site Anderson model in the linearized DMFT. The same model was also used in the ED method with the smallest cluster size $n_s=2$ and obtained the self-consistent solution which does not show the MIT up to $U=100t$. The difference between two methods is the self-consistent procedure. In the linearized DMFT, we focus only on the lowest-energy poles and determine them self-consistently. On the other hand, in the ED method, all of the poles are determined self-consistently so as to satisfy the self-consistency condition as close as possible even in the case with $n_s=2$. The $n_s=2$ system is insufficient to describe the high-energy poles and fails to obtain the MIT within the ED method, while it is sufficient to describe the low-energy poles to obtain the critical value of the MIT within the linearized DMFT.

In the linearized DMFT, we have assumed that as the MIT is approached the central quasiparticle peak becomes isolated from the upper and lower Hubbard bands and the $p$-band. 
Such an isolated quasiparticle peak occurs for the single-band Hubbard model \cite{Georges1,Bulla1} and also for the two-band Hubbard model \cite{Mutou,Ono1}. 
The phase boundary smoothly connects the Mott-Hubbard type and the charge-transfer type transitions, showing that the central quasiparticle peak also smoothly changes from one type to the other. Even in the case of large $p$- and/or $d$-bandwidth, a finite $t_{pd}$ yields a hybridization gap between the $p$- and the $d$-bands and a central quasiparticle peak is found within the hybridization gap isolated from these bands in the limit of the MIT point.

When $t_{pd}=0$ and $U>U_c \approx 6t_{dd}$, there is a transition from the Mott insulator to a metal at $\Delta_c=2t_{pp}=\frac{W_p}{2}$. 
In this case the $p$ and the $d$ electron states are decoupled. 
There is a correlated $d$-band and a separate free $p$-band, and the transition at $\Delta_c=2t_{pp}$ is simply due to the overlap of these bands. The quasiparticle weight in the $p$-band remains unity as the transition is approached. 
This is quite different from the Mott MIT considered here, which is a many-body transition  where the weight and the width of the quasiparticle peak within the hybridization gap due to $t_{pd}$ decrease to zero. 
The value of $\Delta_c$ for this transition, given by eq.(\ref{DC}) in the limit $t_{pd}\to 0$, differs from the value $\Delta_c=2t_{pp}$ for the simple overlap transition, showing that  $t_{pd}=0$ is a singular point of the two-band Hubbard model.

Zaanen, Sawatzky and Allen previously obtained a similar phase diagram of the Mott MIT by using the local impurity approximation (LIA) \cite{Zaanen}. They mapped the lattice model onto the impurity Anderson model where the hybridization function is assumed and is not determined self-consistently in contrast to the DMFT. In the DMFT, the self-consistency condition plays an important role to take into account of the translational symmetry. By neglecting the translational symmetry, the LIA fails to describe precisely the metallic state and the MIT phase boundary although it gives reasonable description for the insulating state when the gap is large \cite{Gap}. 
For the case with small $p$-bandwidth $W_p=4t_{pp}$, the LIA predicts a negative value of $\Delta_c$ in the charge-transfer type \cite{Mizokawa} in contrast to the DMFT where $\Delta_c$ is always positive. 
As mentioned in Section~\ref{sec:3.3}, $\Delta_c$ is roughly given by an energy balance between the charge-transfer energy and the kinetic energy. Even in the case with $t_{pp}=t_{dd}=0$, the kinetic energy due to $t_{pd}$, which is underestimated in the LIA, gives a positive value of $\Delta_c$ in the DMFT.

Within the linearized DMFT, we have determined the chemical potential by using the condition that the hybridization function has a minimum instead of fixing the electron density to be unity. The chemical potential thus obtained is equivalent to the exact result in the single-band Hubbard model on the Bethe lattice. It also agrees well with the result from the ED method in the two-band Hubbard model. This method is applicable for the single-band Hubbard model without particle-hole symmetry such as the fcc-type lattice in $d=\infty$ \cite{Ulmke}. It is also interesting to apply this method to more complicated models, {\it e.g.}, the model with the orbital degeneracy and the Hund rule coupling, where the numerical method of the full DMFT becomes rather difficult.

\section{Acknowledgments}

We would like to thank the Isaac Newton Institute for Mathematical Sciences for hospitality while part of this work was done. 
One of us (Y\=O) was supported by the Grant-in-Aid for Scientific Research from the Ministry of Education, Science, Sports and Culture, and also by CREST (Core Research for Evolutional Science and Technology) of Japan Science and Technology Corporation (JST). 
One of us (RB) was supported by the Deutsche Forschungsgemeinschaft, through the Sonderforschungsbereich 484. 
We also acknowledge the support of the EPSRC grant (GR/M44262) for one of us (ACH).

\appendix
\section{\hspace{-1mm}ppendix A}
\label{Apex:1}

Here we discuss the two-site Anderson model eq.(\ref{2SITE}) \cite{Hewson}. We assume that the conduction level is between the atomic $f$-level and the upper Hubbard level: 
$
\varepsilon_f<\varepsilon_c < \varepsilon_f+U. 
$

The one electron eigenstates 
\begin{eqnarray} 
|E_{\pm}\rangle =  \alpha_{\pm}f_\sigma^+|0\rangle 
                 + \beta_{\pm}c_\sigma^+|0\rangle ,
\end{eqnarray} 
correspond to the eigenenergies
\begin{eqnarray} 
E_{\pm} =  \frac{1}{2}\left(\varepsilon_c+\varepsilon_f
                  \pm\sqrt{(\varepsilon_c-\varepsilon_f)^2+4V^2}\right). 
         \label{EPM}
\end{eqnarray} 
For the small hybridization 
$V^2 \ll \varepsilon_c-\varepsilon_f$, 
eq.(\ref{EPM}) is simplified as
\begin{eqnarray} 
E_{+} =  \varepsilon_c +\frac{V^2}{\varepsilon_c-\varepsilon_f}, \ \ \ 
E_{-} =  \varepsilon_f -\frac{V^2}{\varepsilon_c-\varepsilon_f}, 
\end{eqnarray}
to leading order in $V^2$, with the corresponding eigenstates 
\begin{eqnarray} 
|E_{+}\rangle =  \alpha\left(\frac{V}{\varepsilon_c-\varepsilon_f} f_\sigma^+
                        + c_\sigma^+  \right) |0\rangle,
         \\
|E_{-}\rangle =  \alpha\left(f_\sigma^+ 
     - \frac{V}{\varepsilon_c-\varepsilon_f} c_\sigma^+  \right) |0\rangle,
\end{eqnarray} 
with
$
\alpha = 1-\frac{V^2}{2(\varepsilon_c-\varepsilon_f)^2}. 
$
Similarly, we obtain the three electron (one hole) eigenenergies 
\begin{eqnarray} 
\bar{E}_{\pm} =  \frac{1}{2}\left(3\varepsilon_c+3\varepsilon_f+U
                \pm\sqrt{(\varepsilon_f+U-\varepsilon_c)^2+4V^2}\right). 
         \label{EPMB}
\end{eqnarray} 
For the small hybridization 
$V^2 \ll \varepsilon_f+U-\varepsilon_c$, 
eq.(\ref{EPMB}) is simplified as
\begin{eqnarray} 
\bar{E}_{+} &=&  \varepsilon_c +2\varepsilon_f+U 
               +\frac{V^2}{\varepsilon_f+U-\varepsilon_c}, \\ 
\bar{E}_{-} &=&  2\varepsilon_c+\varepsilon_f 
               -\frac{V^2}{\varepsilon_f+U-\varepsilon_c}, 
\end{eqnarray}
to leading order in $V^2$. The corresponding eigenstates are 
\begin{eqnarray} 
|\bar{E}_{+}\rangle =  \bar{\alpha}\left(\frac{V}{\varepsilon_f+U-\varepsilon_c}                   f_\sigma + c_\sigma  \right) |4\rangle,
         \\
|\bar{E}_{-}\rangle =  \bar{\alpha}\left(f_\sigma  
   - \frac{V}{\varepsilon_f+U-\varepsilon_c} c_\sigma  \right) |4\rangle,
\end{eqnarray} 
with 
$
\bar{\alpha} = 1-\frac{V^2}{2(\varepsilon_f+U-\varepsilon_c)^2} 
$
and 
$
|4\rangle=f_\uparrow^+ f_\downarrow^+ c_\uparrow^+ c_\downarrow^+|0\rangle.
$

The two electron states can be classified as singlets or triplets. In the triplet state, the spatial part of the wavefunction is antisymmetric and the interaction $U$ plays no role. Then the total energy of the triplet state is given by 
$
E_{+} + E_{-} = \varepsilon_c+\varepsilon_f. 
$
There are three possible singlet states which can be written by the linear combination of the states,
\begin{eqnarray} 
|\phi_1\rangle &=& \frac{1}{\sqrt{2}} (c_\uparrow^+ f_\downarrow^+ 
                      - c_\downarrow^+ f_\uparrow^+ )|0\rangle,  \\
|\phi_2\rangle &=& c_\uparrow^+ c_\downarrow^+ |0\rangle,  \\
|\phi_3\rangle &=& f_\uparrow^+ f_\downarrow^+ |0\rangle. 
\end{eqnarray} 
The eigenenergies are given by the solutions of the equation,
\begin{eqnarray} 
\left|
\begin{array}{ccc}
E-\varepsilon_c-\varepsilon_f \  & -\sqrt{2}V & -\sqrt{2}V \\
 -\sqrt{2}V & E-2\varepsilon_c & 0 \\
 -\sqrt{2}V & 0 & E-2\varepsilon_f -U \\
\end{array}
\right|
=0.
\end{eqnarray}
To leading order in $V^2$, the eigenenergies are 
\begin{eqnarray} 
E_{1} &=&  \varepsilon_c +\varepsilon_f 
               -\frac{2V^2}{\varepsilon_c-\varepsilon_f} 
               -\frac{2V^2}{\varepsilon_f+U-\varepsilon_c}, \\ 
E_{2} &=&  2\varepsilon_c 
               +\frac{2V^2}{\varepsilon_c-\varepsilon_f},  \\
E_{3} &=&  2\varepsilon_f+U 
               +\frac{2V^2}{\varepsilon_f+U-\varepsilon_c}, 
\end{eqnarray}
and the corresponding eigenstates are 
\begin{eqnarray} 
|E_{1}\rangle &=& \alpha_1\left(|\phi_1\rangle 
           -\frac{\sqrt{2}V}{\varepsilon_c-\varepsilon_f} |\phi_2\rangle
           -\frac{\sqrt{2}V}{\varepsilon_f+U-\varepsilon_c} |\phi_3\rangle
            \right) , \\
|E_{2}\rangle &=& \alpha_2\left(
            \frac{\sqrt{2}V}{\varepsilon_c-\varepsilon_f} |\phi_1\rangle
           +|\phi_2\rangle
            \right) , \\
|E_{3}\rangle &=& \alpha_3\left(
            \frac{\sqrt{2}V}{\varepsilon_f+U-\varepsilon_c} |\phi_1\rangle
           +|\phi_3\rangle
            \right) , 
\end{eqnarray}
with 
$
\alpha_1 = 1-\frac{V^2}{(\varepsilon_c-\varepsilon_f)^2}
            -\frac{V^2}{(\varepsilon_f+U-\varepsilon_c)^2}, 
$
$
\alpha_2 = 1-\frac{V^2}{(\varepsilon_c-\varepsilon_f)^2}, 
$
and
$
\alpha_3 = 1-\frac{V^2}{(\varepsilon_f+U-\varepsilon_c)^2}. 
$
In our situation with 
$
\varepsilon_f<\varepsilon_c < \varepsilon_f+U, 
$
we find the singlet ground state, $|E_{1}\rangle$, with an energy gain of 
$
 \frac{2V^2}{\varepsilon_c-\varepsilon_f} 
+\frac{2V^2}{\varepsilon_f+U-\varepsilon_c} 
$
due to the hybridization. This is equivalent to the energy gain $2J$ in the s-d model with the Kondo coupling $J$ corresponding to the Schrieffer-Wolff transformation for 
$
V^2 \ll \varepsilon_c-\varepsilon_f 
$
and 
$
V^2 \ll \varepsilon_f+U-\varepsilon_c. 
$ 

Now we calculate the $f$-electron Green's function of this model. When a $f,\uparrow$ electron is removed from the ground state $|E_{1}\rangle$, there are two possible final states: $|E_{+}\rangle$ and $|E_{-}\rangle$. Correspondingly, there are two possible single-hole excitations with excitation energies, 
\begin{eqnarray} 
E_+-E_1=-\varepsilon_f+\frac{3V^2}{\varepsilon_c-\varepsilon_f}
                      +\frac{2V^2}{\varepsilon_f+U-\varepsilon_c}
        \equiv -\varepsilon_1, \label{E1} \\
E_--E_1=-\varepsilon_c+\frac{V^2}{\varepsilon_c-\varepsilon_f}
                      +\frac{2V^2}{\varepsilon_f+U-\varepsilon_c}
        \equiv -\varepsilon_2,  
\end{eqnarray}
to leading order in $V^2$. The matrix elements for these transitions are
\begin{eqnarray} 
\langle E_+|f_\uparrow|E_1\rangle &=& \frac{\alpha \alpha_1}{\sqrt{2}} \left(
              1  -\frac{2V^2}
             {(\varepsilon_c-\varepsilon_f)(\varepsilon_f+U-\varepsilon_c)}
            \right), \nonumber \\
\langle E_-|f_\uparrow|E_1\rangle &=& \frac{\alpha \alpha_1}{\sqrt{2}} \left(
             -\frac{V}{\varepsilon_c-\varepsilon_f}
             -\frac{2V}{\varepsilon_f+U-\varepsilon_c}
            \right), \nonumber
\end{eqnarray}
which yield the transition probabilities: 
\begin{eqnarray} 
|\langle E_+|f_\uparrow|E_1\rangle|^2 &=& \frac{1}{2} 
             -\frac{3V^2}{2(\varepsilon_c-\varepsilon_f)^2}
             -\frac{V^2}{(\varepsilon_f+U-\varepsilon_c)^2}  
             \nonumber \\
             & & -\frac{2V^2}
             {(\varepsilon_c-\varepsilon_f)(\varepsilon_f+U-\varepsilon_c)}
             \nonumber \\
             &\equiv& w_1,  \\
|\langle E_-|f_\uparrow|E_1\rangle|^2 &=& 
              \frac{V^2}{2(\varepsilon_c-\varepsilon_f)^2}
             +\frac{2V^2}{(\varepsilon_f+U-\varepsilon_c)^2}  
             \nonumber \\
             & & +\frac{2V^2}
             {(\varepsilon_c-\varepsilon_f)(\varepsilon_f+U-\varepsilon_c)}
             \nonumber \\
             &\equiv& w_2 , 
\end{eqnarray}
to leading order in $V^2$. 

When a $f,\uparrow$ electron is added to the ground state $|E_{1}\rangle$, possible final states are $|\bar{E}_{-}\rangle$ and $|\bar{E}_{+}\rangle$. Correspondingly, there are two possible single-particle excitations with excitation energies, 
\begin{eqnarray} 
\bar{E}_--E_1&=&\varepsilon_c+\frac{2V^2}{\varepsilon_c-\varepsilon_f}
                      +\frac{V^2}{\varepsilon_f+U-\varepsilon_c}
        \equiv \varepsilon_3,  \\
\bar{E}_+-E_1&=&\varepsilon_f+U+\frac{2V^2}{\varepsilon_c-\varepsilon_f}
                      +\frac{3V^2}{\varepsilon_f+U-\varepsilon_c} \nonumber \\
        &\equiv& \varepsilon_4,  
\end{eqnarray}
to leading order in $V^2$. The matrix elements for these transitions are
\begin{eqnarray} 
\langle \bar{E}_-|f_\uparrow^+|E_1\rangle &=& 
              \frac{\bar{\alpha} \alpha_1}{\sqrt{2}} \left(
              \frac{2V}{\varepsilon_c-\varepsilon_f}
             +\frac{V}{\varepsilon_f+U-\varepsilon_c}
            \right), \nonumber \\
\langle \bar{E}_+|f_\uparrow^+|E_1\rangle &=& 
             \frac{\bar{\alpha} \alpha_1}{\sqrt{2}} \left(
             1  -\frac{2V^2}
             {(\varepsilon_c-\varepsilon_f)(\varepsilon_f+U-\varepsilon_c)}
            \right), \nonumber
\end{eqnarray}
which yield the transition probabilities: 
\begin{eqnarray} 
|\langle \bar{E}_-|f_\uparrow^+|E_1\rangle|^2 &=& 
              \frac{2V^2}{(\varepsilon_c-\varepsilon_f)^2}
             +\frac{V^2}{2(\varepsilon_f+U-\varepsilon_c)^2}  
             \nonumber \\
             & & +\frac{2V^2}
             {(\varepsilon_c-\varepsilon_f)(\varepsilon_f+U-\varepsilon_c)}
             \nonumber \\
             &\equiv& w_3 ,  \\
|\langle \bar{E}_+|f_\uparrow^+|E_1\rangle|^2 &=& \frac{1}{2} 
             -\frac{V^2}{(\varepsilon_c-\varepsilon_f)^2}
             -\frac{3V^2}{2(\varepsilon_f+U-\varepsilon_c)^2}  
             \nonumber \\
             & & -\frac{2V^2}
             {(\varepsilon_c-\varepsilon_f)(\varepsilon_f+U-\varepsilon_c)}
             \nonumber \\
             &\equiv& w_4, \label{W4}
\end{eqnarray}
to leading order in $V^2$. 

From eqs.(\ref{E1})-(\ref{W4}), we obtain the $f$-electron Green's function which has four poles:
\begin{eqnarray} 
G_\sigma(z) = \sum_{i=1}^4 \frac{w_i}{z-\varepsilon_i}. 
\end{eqnarray}
In the limit $V\to 0$, high-energy poles at $\varepsilon_1 \approx \varepsilon_f$ and $\varepsilon_4 \approx \varepsilon_f+U$ have large residues $w_1 \approx w_2 \approx \frac12$, while low-energy poles merge together at $\varepsilon_2 \approx \varepsilon_3 \approx 0$ with small total residue $w \equiv w_2+w_3$: 
\begin{eqnarray} 
w=            \frac{5V^2}{2(\varepsilon_c-\varepsilon_f)^2}
             +\frac{5V^2}{2(\varepsilon_f+U-\varepsilon_c)^2}  
             \nonumber \\
             +\frac{4V^2}
             {(\varepsilon_c-\varepsilon_f)(\varepsilon_f+U-\varepsilon_c)}.
             \label{W23}
\end{eqnarray}



\begin{thebibliography}{}

\bibitem{Georges1}
For a review, see
A. Georges, G. Kotliar, W. Krauth, M. J. Rozenberg, Rev. Mod. Phys.
\textbf{68}, 13 (1996).

\bibitem{Pruschke}
Th. Pruschke, M. Jarrell, J.K. Freericks, Adv. Phys. \textbf{44}, 187 (1995).

\bibitem{Jarrell}
M. Jarrell, Phys. Rev. Lett. \textbf{69}, 168 (1992).

\bibitem{Caffarel}
M. Caffarel, W. Krauth, Phys. Rev. Lett. \textbf{72}, 1545 (1994). 

\bibitem{Sakai}
O. Sakai, Y. Kuramoto, Solid State Commun. \textbf{89}, 307 (1994).

\bibitem{Bulla}
R. Bulla, A.C. Hewson, Th. Pruschke, J. Phys. Cond. Mat. \textbf{10}, 8365 (1998).

\bibitem{Metzner}
W. Metzner, D. Vollhardt,  Phys. Rev. Lett. \textbf{62}, 324 (1989).

\bibitem{Rozenberg}
M. J. Rozenberg, R. Chitra, G. Kotliar, Phys. Rev. Lett. \textbf{83}, 3498 (1999). 

\bibitem{Schlipf}
J. Schlipf,  M. Jarrell, P.G.J. van Dongen, N. Bl\"umer, S. Kehrein, Th. Pruschke, D. Vollhardt, Phys. Rev. Lett. \textbf{82}, 4890 (1999);
Although they claimed that a coexistence disappears in this paper, the recent result \cite{Blumer} shows the coexistence that agrees with ref. \cite{Rozenberg}. 

\bibitem{Blumer}
N. Bl\"umer, R. Bulla, M. Jarrell, P.G.J. van Dongen, D. Vollhardt, A Newton Institute Workshop on Strongly Correlated Electron Systems - Novel Physics and New Materials, unpublished.

\bibitem{Krauth}
W. Krauth, Phys. Rev. B \textbf{62}, 6860 (2000). 

\bibitem{Bulla4}
R. Bulla, T.A. Costi, D. Vollhardt, cond-mat/0012329.

\bibitem{Bulla1}
R. Bulla, Phys. Rev. Lett. \textbf{83}, 136 (1999). 

\bibitem{Fujimori}
A. Fujimori, F. Minami, S. Sugano, Phys. Rev. B \textbf{29}, 5225 (1984).

\bibitem{Zaanen}
J. Zaanen, G.A. Sawatzky, J.W. Allen,  Phys. Rev. Lett. \textbf{55}, 418 (1985).
\bibitem{Georges2}
A. Georges, G. Kotliar, W. Krauth, Z. Phys. \textbf{B 92}, 313 (1993).

\bibitem{Mutou}
 T. Mutou, H. Takahashi, D. S. Hirashima, J. Phys. Soc. Jpn. \textbf{66}, 2781 (1997).

\bibitem{Ono1}
Y. \=Ono, T. A. Tsuruta, Matsuura, Y. Kuroda, Physica B \textbf{281-282}, 410 (2000).

\bibitem{Ono2}
Y. \=Ono, K. Sano, Proceedings of CREST International Workshop: J. Phys. Chem. Solids \textbf{62}, 285 (2001) .

\bibitem{Bulla2} R. Bulla, (1998); unpublished.

\bibitem{Bulla3}
R. Bulla, M. Potthoff, Eur. Phys. J. B \textbf{13}, 257 (2000).

\bibitem{Bethe}
This simplification is not essential. The general case is discussed in ref.\cite{Bulla2}.

\bibitem{Fermi}
The Fermi level for the impurity Anderson model is set to be zero.

\bibitem{Hewson}
A. C. Hewson, \textit{The Kondo Problem to Heavy Fermions} (Cambridge Univ. Press, 1993).

\bibitem{Ono3}
Y. \=Ono (in preparation).

\bibitem{Shimizu}
Y. Shimizu, O. Sakai, \textit{Computational Physics as a New Frontier in Condensed Matter Research}, edited by H. Takayama {\it et al.} (The Phys. Soc. Jap., 1995), p. 42.

\bibitem{Noack}
R. M. Noack, F. Gebhard, Phys. Rev. Lett. \textbf{82}, 1915 (1999). 

\bibitem{Hole}
This situation corresponds to the hole picture in the real $3d$ transition-metal compounds. However we call electron instead of hole for simplicity. 


\bibitem{Approx}
In fact, $G_p(z)$ has a continuum near $\bar\varepsilon_p$. But the residue $w_p$ is roughly independent of the details of $G_p(z)$ near $\bar\varepsilon_p$. Furthermore, in the absence of $t_{pp}$, $G_p(z)$ has no continuum and we obtain eq.(\ref{WP}) directly without the assumption eq.(\ref{GP}). 

\bibitem{Energy}
For $t_{pd}\gg t_{pp}, t_{dd}$, eqs.(\ref{F1},\ref{F2}) with eq.(\ref{F}) yields the equation $\Delta \approx \frac{4t_{pd}^2}{\Delta}(\sqrt{\frac52} -\frac12)$ which means an energy balance. 

\bibitem{Mizokawa}
T. Mizokawa, A. Fujimori, H. Namatame, K. Akeyama, N. Kosugi, Phys. Rev. B \textbf{49}, 7193 (1994).


\bibitem{Gap}
For example, they obtained a MIT phase boundary where the gap is $0.5t_{pd}$. This is an effect of neglecting translational symmetry. 

\bibitem{Ulmke}
M. Ulmke, Eur. Phys. J. B \textbf{1}, 301 (1998); Ferromagnetism was found away from half-filling. The Mott MIT of this fcc-type lattice was not considered here. 





\end{thebibliography}
\end{document}